\documentclass[num-refs]{wiley-article}
\usepackage[utf8]{inputenc}
\usepackage{lipsum}
\usepackage{booktabs}
\PassOptionsToPackage{hyphens}{url}
\usepackage{hyperref}
\usepackage{xcolor}
\hypersetup{
    colorlinks,
    linkcolor={red!50!black},
    citecolor={blue!50!black},
    urlcolor={black}
}
\newcommand*{\restartrowcolors}{%
  \ifhmode\unskip\fi
  \vadjust{%
    \global\rownum=0 %
  }%
}
\usepackage{subcaption}
\usepackage{textcomp}

\usepackage{listings}
\lstdefinestyle{customcpp}{%
  belowcaptionskip=1\baselineskip,
  breaklines=true,
  xleftmargin=\parindent,
  language=C++,
  showstringspaces=false,
  basicstyle=\linespread{0.4}\small\ttfamily,
  keywordstyle=\bfseries\color{green!40!black},
  numberstyle=\tiny,
  commentstyle=\itshape\color{purple!40!black},
  identifierstyle=\bfseries\color{black},
  stringstyle=\color{red},
  emph={int,char,double,float,unsigned},
  emphstyle=\color{blue},
  morekeywords={uint64_t,uint32_t,__m256i,__m128i,UINT64_C},
}
\definecolor{light-gray}{gray}{0.95}

\usepackage{todonotes}
\usepackage{microtype}
\usepackage{CJKutf8}
\usepackage[per-mode=symbol]{siunitx}
\usepackage{pgfplots}
\usepackage{tikz}
\usepackage{tikzscale}

\usepackage{glossaries}

\makeglossaries

\newglossaryentry{scheme_start_state}{
    name=Scheme Start State,
    description={
        Validates the initial character and sets the state to Scheme State or returns failure.
    }
}

\newglossaryentry{scheme_state}{
    name=Scheme State,
    description={
        Transforms every character until \asciicharacter{:} to lowercase and updates the scheme of the URL.
        Depending on the scheme, it sets the state to File State, Special Relative, or Authority
        State, Path or Authority State, or Opaque State. If no \asciicharacter{:} is found, sets the state to
        No Scheme State.
    }
}

\newglossaryentry{no_scheme_state}{
    name=No Scheme State,
    description={
        Depending on the base URL and the having an opaque path returns failure. Sets the state
        to Fragment State if \asciicharacter{\#} is matched, or depending on the base sets the state to Relative State
        or File State.  
    }
}

\newglossaryentry{special_relative_or_authority_state}{
    name=Special Relative or Authority State,
    description={
        Depending on the current character being \asciicharacter{/} sets the state to Special Authority Ignore Slashes
        State or Relative State.
    }
}

\newglossaryentry{path_or_authority_state}{
    name=Path or Authority State,
    description={
        Sets the state to Authority State or Path state if \asciicharacter{/} is seen.
    }
}

\newglossaryentry{relative_state}{
    name=Relative State,
    description={
        Depending on the current iterated character sets the state to Relative Slash State, Query State
        or the Fragment State.
    }
}

\newglossaryentry{relative_slash_state}{
    name=Relative Slash State,
    description={
        Depending on the scheme being special, sets the state to Special Authority Ignore Slashes State
        or Authority State.
    }
}

\newglossaryentry{special_authority_slashes_state}{
    name=Special Authority Slashes State,
    description={
        Sets the state to Special Authority Ignore Slashes State. Depending on the input starting 
        with \asciicharacter{/} character, increases or decreases the pointer to the current iterated character.
    }
}

\newglossaryentry{special_authority_ignore_slashes_state}{
    name=Special Authority Ignore Slashes State,
    description={
        Ignores each character of the input until \asciicharacter{/} or \asciicharacter{\textbackslash} characters are seen and sets the
        state to Authority State.
    }
}

\newglossaryentry{authority_state}{
    name=Authority State,
    description={
        Parses validates and encodes username until \asciicharacter{:} character is seen, or password until \asciicharacter{@}
        character is seen. Later, it sets the state to Host State.
    }
}

\newglossaryentry{host_state}{
    name=Host State,
    description={
        Parses each character until iterating through \asciicharacter{/}, \asciicharacter{\textbackslash}, \texttt\asciicharacter{?} or \texttt\asciicharacter{\#} characters is
        consumed. Depending on the buffered value, parses the text into an IPv4, Ipv6, or plain domain.
        Runs Domain To ASCII algorithm defined in the Unicode Specification and set the state to 
        Port State, Query State, or Fragment State according to corresponding characters.
    }
}

\newglossaryentry{port_state}{
    name=Port State,
    description={
        Validates 16-bit unsigned integer representation of the port and sets the state to Path Start State.
    }
}

\newglossaryentry{file_state}{
    name=File State,
    description={
        Sets the scheme to file and host to an empty string and sets the state to either Fragment State,
        Query State or Path State depending on the iterated character type.
    }
}

\newglossaryentry{file_slash_state}{
    name=File Slash State,
    description={
        Sets the state to File Host State or Path State depending on iterating through \asciicharacter{/} or 
        \asciicharacter{\textbackslash} characters.
    }
}

\newglossaryentry{file_host_state}{
    name=File Host State,
    description={\todo{Add this}}
}

\newglossaryentry{path_start_state}{
    name=Path Start State,
    description={\todo{Add this}}
}

\newglossaryentry{path_state}{
    name=Path State,
    description={\todo{Add this}}
}

\newglossaryentry{opaque_path_state}{
    name=Opaque Path State,
    description={
        Acts as an intermediary for setting the state to Query State, Fragment State, or appending
        the iterated character to the URL's path depending on the type of iterated character.
    }
}

\newglossaryentry{query_state}{
    name=Query State,
    description={
        Parses and percent-encode each character until the EOF code point or \asciicharacter{\#} character
        is seen.
    }
}

\newglossaryentry{fragment_state}{
    name=Fragment State,
    description={
        Parses and percent-encode each character until the EOF code point is seen.
    }
}

\usetikzlibrary{automata,positioning,arrows,quotes}

\definecolor{bblue}{HTML}{4F81BD}
\definecolor{rred}{HTML}{C0504D}
\definecolor{ggreen}{HTML}{9BBB59}
\definecolor{ppurple}{HTML}{9F4C7C}
\definecolor{ggreen}{HTML}{00FF00}

\newcommand{\asciicharacter}[1]{`\texttt{#1}'}

\papertype{Research Article}

\title{Parsing Millions of URLs per Second}
\author[1]{Yagiz Nizipli}
\author[1\authfn{1}]{Daniel Lemire}

\affil[1]{Data Science Research Center, Universit\'e du Qu\'ebec (TELUQ), Montreal, Quebec, H2S 3L5, Canada}

\corraddress{Daniel Lemire, Data Science Research Center, Universit\'e du Qu\'ebec (TELUQ), Montreal, Quebec, H2S 3L5, Canada}
\corremail{daniel.lemire@teluq.ca}

\fundinginfo{Natural Sciences and Engineering Research Council of Canada, Grant Number: RGPIN-2017-03910}

\runningauthor{Yagiz Nizipli and Daniel Lemire}
\begin{document}


\maketitle
\begin{abstract}
URLs are fundamental elements of web applications. 
By applying  vector algorithms,  we built a fast standard-compliant C++ implementation.
Our parser uses three times fewer instructions than competing parsers following the WHATWG standard (e.g., Servo's rust-url) and up to eight times fewer instructions than the popular curl parser.
The  Node.js environment adopted our C++ library. In our tests on realistic data, a recent Node.js version (20.0) with our parser is four to five times faster than the last version with the legacy  URL parser.
\keywords{URL, Text Processing, Vectorization, Performance}
\end{abstract}
\section{Introduction}

A Uniform Resource Locator (URL) is a unique identifier that defines
a resource on the web. A typical URL might provide a protocol, a domain and a path. We offer several URL examples in Table~\ref{table:urlexamples}.

Berners-Lee et al. (2005) defined 
the URL syntax was defined  in the Request for Comments~(RFC)~3986~\cite{rfc3986}. However,
the standard evolved organically through various implementations over the years. To address the growing difference between existing standards
and practice---most notably in web browsers---the Web Hypertext Application Technology Working Group proposed the WHATWG URL Standard in 2012~\cite{whatwgurl}.
Most of the popular web browsers (e.g., from Apple, Mozilla or Google) abide
by the WHATWG URL standard. 
Unfortunately, many standard libraries still fail to follow the WHATWG URL standard: we verified that support is missing in the Java standard library (\texttt{java.net.URL}) as of Java~20,
in the Go standard library (\texttt{net/url}) as of Go~1.21, in PHP (\texttt{parse\_url}) as of PHP~8.1 and in Python's
\texttt{urllib} library as of Python~3.11. Furthermore, popular URL parsers often differ in how they interpret URL strings~\cite{ajmani2022you,reynolds2022equivocal}.

URL parsing consists in taking an input string and identifying the various components while normalizing them as needed. For example, the input string \texttt{\begin{CJK*}{UTF8}{bsmi}http://你好你好.在线\end{CJK*}/./a/../b/./c} should be normalized to the string\begin{equation*}
    \texttt{https://xn--6qqa088eba.xn--3ds/b/c}
\end{equation*}  where \texttt{https:} represent the protocol, \texttt{xn--6qqa088eba.xn--3ds} is the host, and \texttt{/b/c} is the path.
We may also need to parse a URL string relative to another string. For
example, given the base string \texttt{http://example.org/foo/bar}, the
relative string \texttt{http:/example.com/} leads to the final URL \texttt{http://example.org/example.com/}. We should also be able to modify the various components of a URL (protocol, host, username, etc.). To illustrate the complexity, our C++ software library implementing the WHATWG URL standard---and little else---has approximately \num{20000}~lines of code.

The WHATWG URL standard follows  the robustness principle (Postel's law): 
\emph{be conservative in what you send, be liberal in what you accept}.
Parsing URLs using the WHATWG URL standard can be more challenging than using the earlier standard (RFC~3986). For example,
consider the string \texttt{https://\textcolor{blue}{\textbackslash{}t}lemire.me/en/} where \texttt{\textcolor{blue}{\textbackslash{}t}} is the tabulation character. The WHATWG URL standard requires us to ignore the tabulation characters. A conventional URL parser following RFC~3986 (e.g., curl\footnote{Curl stands for \emph{command line tool and library 
for transferring data with URLs} and it is sometimes capitalized as \emph{cURL} though the official documentation and website use a lowercase name: \emph{curl}. })  would reject such a string.

There are many components that impact the performance of a web application but
URL parsing is practically always required. 
URL parsing is relatively expensive.
Parsing a single URL may take \SI{4}{\micro\second} on average in a system like Node.js.
In our tests, the popular curl library can parse about half a million URLs per second, yet
a fast C++ number parser---converting ASCII number strings to binary floating-point numbers---can process more than 50~million numbers per second~\cite{lemire2021number}. Thus we can parse almost 100 floating-point numbers in the time it takes curl to parse a single URL\@.

We think that popular systems such as Node.js should be able to parse several million URLs per second on modern systems without sacrificing correctness or safety.
We present our work on  the efficient implementation  of the current WHATWG specification.
Our implementation is freely available.\footnote{\url{https://www.github.com/ada-url/ada}}
We provide benchmarks and comparisons with other fast and popular URL parsers
in C, C++, and Rust, 
whether they follow RFC~3986~\cite{rfc3986} (curl and Boost.URL) 
or WHATWG URL (Servo rust-url).
We  review various strategies that are 
efficient when parsing strings.

Our work has been integrated into the popular Node.js JavaScript runtime environment over several versions, concluding with a final integration in Node.js version~20.
We are therefore able to run JavaScript benchmarks before the inclusion of 
our fast parser (e.g., Node.js version~18) and after its complete integration
(e.g., Node.js version~20). Though many factors contribute to improved performance, 
we estimate that the large performance gains in URL parsing are mostly the result of our work.



\begin{table}
\caption{URL examples. \label{table:urlexamples}}
\begin{tabular} {lp{0.7\columnwidth}}
\toprule
Long URLs & \url{http://nodejs.org:89/docs/latest/api/foo/bar/qua/13949281/0f28b//5d49/b3020/url.html#test?payload1=true&payload2=false&test=1&benchmark=3&foo=38.38.011.293&bar=1234834910480&test=19299&3992&key=f5c65e1e98fe07e648249ad41e1cfdb0} \\
Short URLs & \url{https://nodejs.org/en/blog/} \\
IDN & \begin{CJK*}{UTF8}{bsmi}http://你好你好.在线\end{CJK*}\\
File & \url{file:///foo/bar/test/node.js} \\
Websocket & \url{ws://localhost:9229/f46db715-70df-43ad-a359-7f9949f39868} \\
Authentication & \url{https://user:pass@example.com/path?search=1} \\
JavaScript & \url{javascript:alert("node is awesome");} \\
Percent Encoding & \url{https://\%E4\%BD\%A0/foo} \\
Pathname with dots & \url{https://example.org/./a/../b/./c} \\
\bottomrule
\end{tabular}
\end{table}

\section{Related Work}


Much of the academic research regarding URLs relates to security issues. For example, 
Ajmani et al.~\cite{ajmani2022you} as well as 
Reynolds et al.~\cite{reynolds2022equivocal}
test a wide range of popular URL parsers: they find many differences and discuss the security implications of these differences. 
In our work, we sought to provide complete and rigorous support to the WHATWG URL specification.

To our knowledge, there is no related work on the production of high-performance URL parsers.  
However, there is  related work regarding the high-performance parsing of web formats. 
Park et al.~\cite{park2016concurrent} show that we can improve the performance of web applications by parsing JavaScript concurrently. XML  parsing has received much attention: e.g., Van Engelen proposes fast XML parsing with  deterministic finite state automata~\cite{van2004constructing}, 
Kostoulas et al.\ achieve higher XML parsing speed by avoiding unnecessary data copying and transformation~\cite{Kostoulas:2006:XSI:1135777.1135796}, Cameron et al.\ show that we can parse XML faster using SIMD instructions~\cite{Cameron:2008:HPX:1463788.1463811}. There is also much work regarding JSON parsing: e.g., Langdale and Lemire show that we can parse gigabytes of JSON per second using branchless routines and vectorization~\cite{langdale2019parsing}.
Binary data is commonly published online using 
segments of base64 code: we can greatly accelerate
the coding and decoding of these segments~\cite{mula2018faster,mula2020base64}.






\section{Parsing URL Strings}


A URL string consists of many  substrings. We refer to these substrings as components.
Once normalized, URL strings are ASCII but a parser may receive a Unicode string.

A URL string typically begins with a protocol (or \emph{scheme}) string: e.g., the string  \texttt{http} in \texttt{http://google.com}. The WHATWG URL specification recognizes \emph{special} protocols that are subject to different constraints:  \texttt{ftp}, \texttt{file}, \texttt{http}, \texttt{https}, \texttt{ws}, and \texttt{wss}. The protocol string  is terminated by the colon character (\asciicharacter{:}). 

The protocol might be followed by a host. In such cases, the protocol-terminating colon character is followed by two slash characters
\asciicharacter{//}. 
A host might be preceded by \emph{credentials}.
Credentials in a URL define the username with an optional password split with the
\asciicharacter{:} character. E.g., \url{postgresql://username:password@localhost:5432}. To have credentials, a URL string must not have the protocol \texttt{file} and it must have a non-empty host.
A host begins with a hostname string, optionally followed by the colon character (\asciicharacter{:}) and a port number string.
Thus, given the URL string \texttt{data://example.com:8080/pathname?search}, the host is \texttt{example.com:8080} whereas the hostname is \texttt{example.com}.
URL strings with a special protocol must contain a \emph{host}, whereas it is optional for other types of URL strings.
Host names may be domain names, IPv4, or IPv6 addresses.

\begin{itemize}
    \item For non-ASCII domain names, we must follow RFC~2390~\cite{rfc3490} which involves converting Unicode to punycode~\cite{rfc3492} and checking that various rules are satisfied. 

\item The IPv4 address is a 32-bit unsigned integer that identifies a network address. The 
WHATWG  URL specification considers both \texttt{192.168.1.1} and \texttt{192.0x00A80001} as valid and equivalent IPv4 addresses. The normalized URL string is made of four decimal integers (\texttt{192.168.1.1}).

\item The IPv6 address is a 128-bit unsigned integer that identifies a network address. It
is represented as a list of eight 16-bit unsigned integers, also known as IPv6
        pieces. We surround IPv6 addresses by square brackets:  e.g., \texttt{http://[c141:ffff:0:ffff:ffff:ffff:ffff:ffff]}.

\item Port numbers are represented by 16-bit integers with a maximum value of 65536. Special protocol have default ports: e.g., the \texttt{http} protocol has default port 80. 
Default ports are omitted in the normalized string. The \texttt{file} protocol cannot have a port. It is also disallowed to have port without a hostname.
\end{itemize}

It is possible for a URL to have no host (and thus no credentials), in which case the protocol string is not followed by two slashes \asciicharacter{//}: e.g., \texttt{non-spec:/.//p}. The standard distinguishes between an empty host (e.g., \texttt{protocol:///mypath}) and a missing host (e.g., \texttt{protocol:/mypath}).

A URL string may contain a \emph{pathname}
after the protocol and (optional) host.
If there is no host then the pathname is opaque: e.g., 
the URL  \texttt{mailto:john@doe.com} has the opaque pathname 
\texttt{john@doe.com}.
Otherwise a  URL pathname starts with \asciicharacter{/}.
If the host is empty, there might be a sequence of three slash characters: e.g. \texttt{file:///file.txt}.
The pathname is always optional.
If the pathname contains non-ASCII characters, they are \emph{percent encoded}: treating
the characters as UTF-8 bytes, we replace non-ASCII characters with a sequence of \asciicharacter{\%} characters followed by two-character hexadecimal codes.  
For example, the character \texttt{é} is replaced by the sequence
\texttt{\%C3\%A9}.

We may then have a search component 
(also called a \emph{query}). 
A URL search component is represented 
by either null or an ASCII string and 
starts with the character \asciicharacter{?}.
It is usual for the search component 
to content a sequence of key-value pairs separated by the ampersand character 
\asciicharacter{{\&}} and 
linked by the equal sign \asciicharacter{=}: 
\texttt{?a=b\&c=d}.
The search component is percent-encoded as needed.
Similarly, we may have a hash component
(also called a \emph{fragment}). The 
    URL hash is the URL part that starts with the \asciicharacter{\#} character. It may also be percent-encoded.



\section{Fast Parsing}

\begin{figure}[tb!]
\begin{tikzpicture}[
    ->,
    >=stealth',
    shorten >=1pt,
    auto,
    node distance=3cm,
    scale = 0.9,
    transform shape
    ]

  \node[state,initial] (scheme_start) {\scriptsize\begin{tabular}{@{}c@{}}Scheme\\Start\end{tabular}};
  \node[state,accepting,below of=scheme_start] (scheme) {\scriptsize\begin{tabular}{@{}c@{}}Scheme\end{tabular}};
  \node[state,above left of=scheme] (file) {\scriptsize\begin{tabular}{@{}c@{}}File\end{tabular}};
  \node[state,below left of=scheme, align=center] (special_relative_or_authority) {\scriptsize\begin{tabular}{@{}c@{}}Special\\relative or\\Authority\end{tabular}};
  \node[state,below of=scheme] (special_authority_slashes) {\scriptsize\begin{tabular}{@{}c@{}}Special\\Authority\\Slashes\end{tabular}};
  \node[state, below right of=scheme] (path_or_authority) {\scriptsize\begin{tabular}{@{}c@{}}Path or\\Authority\end{tabular}};
  \node[state,right of=path_or_authority] (opaque_path) {\scriptsize\begin{tabular}{@{}c@{}}Opaque\\Path\end{tabular}};
  \node[state,accepting,above right of=scheme] (no_scheme) {\scriptsize\begin{tabular}{@{}c@{}}No\\Scheme\end{tabular}};
  \node[state, right of=no_scheme] (relative) {\scriptsize\begin{tabular}{@{}c@{}}Relative\end{tabular}};
  \node[state,accepting,below right of=no_scheme] (fragment) {\scriptsize\begin{tabular}{@{}c@{}}Fragment\end{tabular}};
  \node[state,below left of=special_authority_slashes] (special_authority_ignore_slashes) {\scriptsize\begin{tabular}{@{}c@{}}Special\\Authority\\Ignore\\Slashes\end{tabular}};
  \node[state,right of=special_authority_ignore_slashes] (authority) {\scriptsize\begin{tabular}{@{}c@{}}Authority\end{tabular}};
  \node[state,accepting,below right of=path_or_authority] (path) {\scriptsize\begin{tabular}{@{}c@{}}Path\end{tabular}};
  \node[state,accepting,right of=opaque_path] (query) {\scriptsize\begin{tabular}{@{}c@{}}Query\end{tabular}};
  
  \node[state,accepting,below of=authority] (host) {\scriptsize\begin{tabular}{@{}c@{}}Host\end{tabular}};
  \node[state,left of=host] (file_host) {\scriptsize\begin{tabular}{@{}c@{}}File\\Host\end{tabular}};
  \node[state,below of=host, yshift=1cm] (path_start) (path_start) {\scriptsize\begin{tabular}{@{}c@{}}Path\\Start\end{tabular}};
  \node[state,accepting,right of=host] (port) {\scriptsize\begin{tabular}{@{}c@{}}Port\end{tabular}};
  
  \node[state,above right of=port] (relative_slash) {\scriptsize\begin{tabular}{@{}c@{}}Relative\\Slash\end{tabular}};
  \node[state] (file_slash) [left of=file, xshift=1cm] {\scriptsize\begin{tabular}{@{}c@{}}File\\Slash\end{tabular}};
  \path (scheme_start) edge                                 node {$ $} (scheme)
        (scheme) edge [loop right]                          node {$ $} (scheme)
        (scheme) edge [bend left]                           node {$ $} (file)
        (scheme) edge                                       node {$ $} (special_relative_or_authority)
        (scheme) edge                                       node {$ $} (special_authority_slashes)
        (scheme) edge                                       node {$ $} (path_or_authority)
        (scheme) edge                                       node {$ $} (opaque_path)
        (scheme) edge                                       node {$ $} (no_scheme)
        (no_scheme) edge                                    node {$ $} (relative)
        (no_scheme) edge                                    node {$ $} (fragment)
        (no_scheme) edge                                    node {$ $} (file)
        (special_relative_or_authority) edge                node {$ $} (special_authority_ignore_slashes)
        (special_relative_or_authority) edge                node {$ $} (relative)
        (path_or_authority) edge                            node {$ $} (authority)
        (path_or_authority) edge                            node {$ $} (path)
        (relative) edge   [bend left]                       node {$ $} (relative_slash)
        (relative) edge                                     node {$ $} (query)
        (relative) edge                                     node {$ $} (fragment)
        (relative) edge   [out=305,in=35]                   node {$ $} (path)
        (relative_slash) edge  [bend left]                  node {$ $} (special_authority_ignore_slashes)
        (relative_slash) edge                               node {$ $} (authority)
        (special_authority_slashes) edge                    node {$ $} (special_authority_ignore_slashes) 
        (special_authority_ignore_slashes) edge             node {$ $} (authority)
        (authority) edge [out=220,in=250,looseness=4]       node {$ $} (authority)
        (authority) edge                                    node {$ $} (host)
        (host) edge                                         node {$ $} (file_host)
        (host) edge                                         node {$ $} (path_start)
        (host) edge  [out=230,in=260,looseness=8]           node {$ $} (host)
        (host) edge                                         node {$ $} (port)
        (port) edge [loop below]                            node {$ $} (port)
        (port) edge                                         node {$ $} (path_start)
        (file) edge                                         node {$ $} (file_slash)
        (file) edge[out=325,in=160]                         node {$ $} (query)
        (file) edge                                         node {$ $} (fragment)
        (file) edge [out=240,in=270,looseness=8]            node {$ $} (file)
        (file_slash) edge [bend right]                      node {$ $} (file_host)
        (file_slash) edge[out=305,in=150]                   node {$ $} (path)
        (file_host) edge                                    node {$ $} (path)
        (file_host) edge                                    node {$ $} (path_start)
        (file_host) edge [loop above]                       node {$ $} (file_host)
        (path_start) edge                                   node {$ $} (path)
        (path_start) edge                                   node {$ $} (query)
        (path_start) edge                                   node {$ $} (fragment)
        (path) edge [out=260,in=290,looseness=8]            node {$ $} (path)
        (path) edge [bend right]                            node {$ $} (query)
        (path) edge                                         node {$ $} (fragment)
        (opaque_path) edge [bend right]                     node {$ $} (fragment)
        (opaque_path) edge  [out=240,in=270,looseness=8]    node {$ $} (opaque_path)
        (query) edge                                        node {$ $} (fragment)
        (query) edge [loop above]                           node {$ $} (query)
        (fragment) edge [loop right]                        node {$ $} (fragment);
\end{tikzpicture}
\caption{URL Parser State Machine\label{fig:url-state-machine}}
\end{figure}
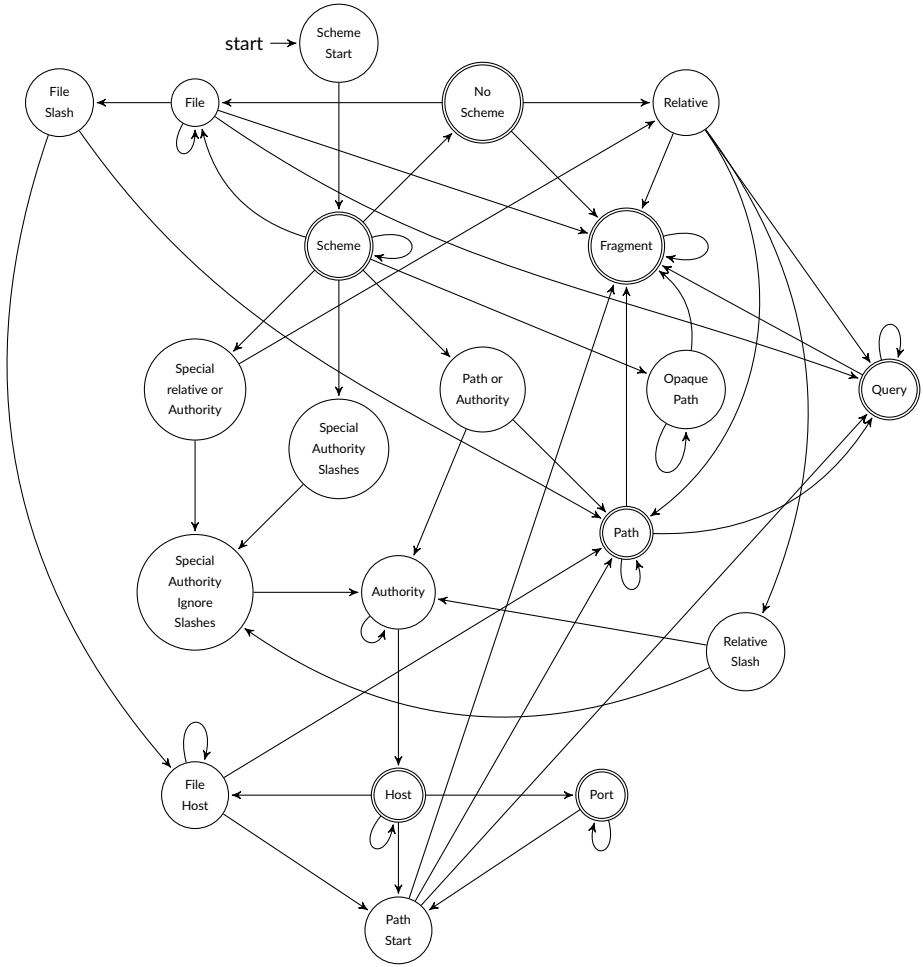

The WHATWG URL standard is specified  as an algorithm following a state-machine. See Fig.~\ref{fig:url-state-machine}. URL parsing begins in the \emph{Scheme
Start} state. The algorithm consumes one character at a time, and changes state according to state-specific rules. In certain
scenarios, the URL state machine reverses the iteration and goes back, resulting in re-iterating the same character more than once. 

We wrote our parser in C++ by initially following the finite-state design. However, the byte-by-byte processing implied by the standard is a poor choice for performance. Thus we adapted the design so that once we enter a state, we fully consume the relevant component of the URL string, as much as possible.

The standard also suggests that each component is parsed into a separate string instance. Though we optionally support this design, our default is to parse into a single string which constitutes the normalized string at the end of the parsing. We call the result an \texttt{url\_aggregator} because the components are aggregated during parsing into a single buffer. Having a single buffer has several performance benefits:
\begin{itemize}
    \item At the beginning of the parsing, we allocate a buffer that has the size of the input string, rounded up to the next power of two. Usually there is no need for further memory allocation or copying. By allocating less memory, we reduce the probability of incurring expensive cache misses.
    \item When querying for string components, or for the normalized string, there is no need to generate and allocate a new string instance. We may simply return an immutable view on the underlying buffer. It is made convenient by the introduction of the \texttt{string\_view} class in C++17 but it is also convenient in other programming languages: Rust has string slices (\texttt{str}), Java has \texttt{CharSequence}, C\# has \texttt{ReadOnlySpan<char>} and so forth.
\end{itemize}

We expect that most components consumed from input URL strings do not need to be modified and they may  be copied as is. We optimized our code for this scenario by integrating tests leading to fast paths. For example, we must remove tabulation and newline characters from input strings, since they are ignored during the processing. However, most input strings do not contain tabulation and newline characters. Thus we use a fast scanning function to verify that there are no such characters. Common processors (Intel, AMD, ARM, POWER) support single-instruction-multiple-data (SIMD) instructions. SIMD instructions operate on several words at once unlike regular instructions. Though different processors support different SIMD instruction sets, there is some common ground.
The 64-bit processors from Intel and AMD (x64) are required to support SSE2~instructions while 64-bit ARM processors (Apple, Qualcomm, etc.) support NEON instructions. 
We can use these instructions through \emph{intrinsic functions} in C and C++: these special functions often provide functionality similar to  a given instruction (e.g., a NEON addition), without using assembly.
Fig.~\ref{lst:tab-new-line-cpp} illustrates one function scan for characters under x64 processors, using SSE2 intrinsic function. The function loads three 16-byte SIMD variables filled with the characters \textbackslash{}r, \textbackslash{}n, and \textbackslash{}t respectively. We start the loop with a SIMD \texttt{running} variable that contains initially only 16~zero bytes. During each iteration, we load up 16~bytes of data from the input, and we execute three comparisons between the newly loaded 16~bytes and each  one with each of the three registers corresponding to \textbackslash{}r, \textbackslash{}n, and \textbackslash{}t respectively. We combine the result with a bitwise-OR operation. If one of the three characters (\textbackslash{}r, \textbackslash{}n, and \textbackslash{}t) appeared in the input, then at least one element from the \texttt{running} register will be non-zero. We have a final iterator for the case where the input does not contain a multiple of 16~bytes: in this case, we copy the last section of input to a 16-byte array on the stack and load 16~bytes from this array. At the end, and only at the end, we check whether one of the element of the \texttt{running} variable is non-zero with the \texttt{pmovmskb} instruction and a branch. Thus our code always consumes the entire input: we proceed in this manner because we expect inputs to almost never contain the characters \textbackslash{}r, \textbackslash{}n, and \textbackslash{}t. We prefer to save instructions and reduce the number of branches in the common case when the three characters are absent, at the expense of more expensive processing when the one of the three characters are present. In this sense, our approach is \emph{optimistic}: we assume that, most times, our input is as expected and we assume that special cases (e.g., the presence of  \textbackslash{}r, \textbackslash{}n, and \textbackslash{}t within the URL string) are rare.
We have also the equivalent function in NEON, as well as a fallback function for other processors. Both SSE2 and NEON instructions are a standard component of the x64 and aarch64 (64-bit ARM) instruction sets. The compiler routinely compiles C++ code to these instructions (SSE2 and NEON) and they are part of the standard libraries.
We detect the target family of processors at compile time. Effectively, the routine compares each input character with the newline and tabulation characters.
When at one such character is found, we use a slow path
where a temporary buffer is allocated. We write a version of the input string to the temporary buffer while omitting the newline and tabulation characters. We find in practice that it is rarely needed.

\begin{figure}
\begin{tabular}{c} 
\begin{lstlisting}[style=customcpp]
bool has_tabs_or_newline(std::string_view user_input) {
  size_t i = 0;
  const __m128i mask1 = _mm_set1_epi8('\r');
  const __m128i mask2 = _mm_set1_epi8('\n');
  const __m128i mask3 = _mm_set1_epi8('\t');
  __m128i running{0};
  for (; i + 15 < user_input.size(); i += 16) {
    __m128i word = _mm_loadu_si128((const __m128i*)(user_input.data() + i));
    running = _mm_or_si128(
        _mm_or_si128(running, _mm_or_si128(_mm_cmpeq_epi8(word, mask1),
          _mm_cmpeq_epi8(word, mask2))),
          _mm_cmpeq_epi8(word, mask3));
  }
  if (i < user_input.size()) {
    uint8_t buffer[16]{};
    memcpy(buffer, user_input.data() + i, user_input.size() - i);
    __m128i word = _mm_loadu_si128((const __m128i*)buffer);
    running = _mm_or_si128(
        _mm_or_si128(running, _mm_or_si128(_mm_cmpeq_epi8(word, mask1),
          _mm_cmpeq_epi8(word, mask2))),
          _mm_cmpeq_epi8(word, mask3));
  }
  return _mm_movemask_epi8(running) != 0;
}
\end{lstlisting}
  \end{tabular}
  \caption{Scan for tabulation and newline characters using SSE2 intrinsic functions. The ARM NEON version is similar.}\label{lst:tab-new-line-cpp}
\end{figure}

Most of the strings begin with a protocol string (e.g., \texttt{file} or \texttt{https}). We must recognize 
a limited set of \emph{special} protocols specified by the WHATWG URL standard. We identify that first occurrence of the colon character \asciicharacter{:} and
seek to recognize quickly the protocol. We expect
most protocol strings to be special in practice: it is uncommon for protocols not to be one of \texttt{http}, \texttt{https}, \texttt{ws}, \texttt{wss}, \texttt{ftp} or \texttt{file}.
We designed a perfect hash function~\cite{10.5555/331120.331202} (see Fig.~\ref{lst:get-scheme-cpp}).
The function first checks whether the string is empty, a special case. If it is not empty, we use as a hash function, twice the length of the string plus the integer value of the first byte of the string. We select only the the least significant three bits of the results, thus generating a value between 0 and 7 inclusively. For valid special protocols, the hash function  returns a value between 0 and 6 inclusively. It is a perfect hash function:  special protocols are mapped to distinct integer values.
We can verify that the string \texttt{http} is mapped to 0, the string \texttt{https} to 2, and so forth. 
We look up the result in a table (\texttt{http}, , \texttt{https}, \texttt{ws}, \texttt{ftp}, \texttt{wss}, \texttt{file}): 
The function compares the input with the content of the table, so no false positive is possible.
Based solely on the length of the protocol string and the first character, we can distinguish any one of the special protocols.
 At several steps during the processing,  the standard requires us to check the protocol. 
 If we  merely store a string value representing the protocol, then we may need to do a string-to-string comparison each time. 
 Instead, for example, we can verify whether we have \texttt{file} protocol by comparing
the protocol type
with the integer value~6: an integer-to-integer comparison may result in a single instructions once compiled unlike a string comparison.

\begin{figure}
\begin{tabular}{c} 
\begin{lstlisting}[style=customcpp]
std::string_view is_special_list[] = {"http", " ",   
   "https", "ws", "ftp",  "wss", "file",  " "};

enum type {
  HTTP = 0, NOT_SPECIAL = 1, HTTPS = 2, WS = 3,
  FTP = 4, WSS = 5, FILE = 6 };

type get_scheme_type(std::string_view scheme)  {
  if (scheme.empty()) { return NOT_SPECIAL; }
  int hash_value = (2 * scheme.size() + (unsigned)(scheme[0])) & 7;
  std::string_view target = is_special_list[hash_value];
  if ((target[0] == scheme[0]) && (target.substr(1) == scheme.substr(1))) {
    return type(hash_value);
  } else { return NOT_SPECIAL; }
}
\end{lstlisting}
  \end{tabular}
  \caption{Analysis of the protocol string}\label{lst:get-scheme-cpp}
\end{figure}

Most URL strings have a host string that must be processed. In the majority of cases, the host string requires no further processing: it is a lower-case ASCII string. We use the function of Fig.~\ref{lst:scan-host-cpp} to identify problematic characteristics. Effectively, it is a stream of table lookups with bitwise OR operations. Each character is viewed as a byte value (between 0 and 256) and the 256-byte table contains values 0, 1, 2 depending on whether the character is a forbidden character (value 1), an upper case letter (value 2) or a valid character (value 0). The result of the function is zero if the input is a lower-case ASCII string, it is 2 if it is an otherwise valid input but with upper case letters. If the result of the function is 1 or 3, then the input contains invalid characters.
Though we could use SIMD instructions for this purpose, the hostnames are relatively short.
When the host string contains non-ASCII lower-case characters, we fall back on a relatively extensive 
normalization process which may include punycode encoding~\cite{rfc3490,rfc3492}. About half of our C++ source code (or \num{10000}~lines) is dedicated to this normalization: thankfully it is rarely needed in practice. 
We also include a fast routine to detect IPv6 when the host string begins with the bracket \asciicharacter{[}. We also check for IPv4 by scanning for digits and the dot character  \asciicharacter{.}. As soon as an IPv6 or IPv4 address is found, we normalize it using a specialized routine.

\begin{figure}
\begin{tabular}{c} 
\begin{lstlisting}[style=customcpp]

static uint8_t is_forbidden_domain_code_point_table_or_upper[] = {
    1, 1, 1, 1, 1, 1, 1, 1, 1, 1, 1, 1, 1, 1, 1, 1, 1, 1, 1, 1, 1, 1, 1, 1,
    1, 1, 1, 1, 1, 1, 1, 1, 1, 0, 0, 1, 0, 1, 0, 0, 0, 0, 0, 0, 0, 0, 0, 1,
    0, 0, 0, 0, 0, 0, 0, 0, 0, 0, 1, 0, 1, 0, 1, 1, 1, 2, 2, 2, 2, 2, 2, 2,
    2, 2, 2, 2, 2, 2, 2, 2, 2, 2, 2, 2, 2, 2, 2, 2, 2, 2, 2, 1, 1, 1, 1, 0,
    0, 0, 0, 0, 0, 0, 0, 0, 0, 0, 0, 0, 0, 0, 0, 0, 0, 0, 0, 0, 0, 0, 0, 0,
    0, 0, 0, 0, 1, 0, 0, 1, 1, 1, 1, 1, 1, 1, 1, 1, 1, 1, 1, 1, 1, 1, 1, 1,
    1, 1, 1, 1, 1, 1, 1, 1, 1, 1, 1, 1, 1, 1, 1, 1, 1, 1, 1, 1, 1, 1, 1, 1,
    1, 1, 1, 1, 1, 1, 1, 1, 1, 1, 1, 1, 1, 1, 1, 1, 1, 1, 1, 1, 1, 1, 1, 1,
    1, 1, 1, 1, 1, 1, 1, 1, 1, 1, 1, 1, 1, 1, 1, 1, 1, 1, 1, 1, 1, 1, 1, 1,
    1, 1, 1, 1, 1, 1, 1, 1, 1, 1, 1, 1, 1, 1, 1, 1, 1, 1, 1, 1, 1, 1, 1, 1,
    1, 1, 1, 1, 1, 1, 1, 1, 1, 1, 1, 1, 1, 1, 1, 1};

bool contains_forbidden_domain_code_point_or_upper(std::string_view input) {
  uint8_t accumulator{};
  for (size_t i = 0; i < input.size(); i++) {
    accumulator |=
        is_forbidden_domain_code_point_table_or_upper[uint8_t(input[i])];
  }
  return accumulator;
}
\end{lstlisting}
  \end{tabular}
  \caption{Function to detect forbidden of upper-case characters in host string}\label{lst:scan-host-cpp}
\end{figure}






We must then process the rest of the URL string, including the path, the search, and the hash substrings. These components may sometimes require percent-encoding. To avoid unnecessary percent-encoding, we search through each substring for the first character that might require percent encoding, when none is found, we can skip percent encoding entirely. Otherwise, we proceed with percent encoding from that character. We classify characters needing percent encoding using fast table lookups.

A challenge with path processing is that while most path strings require little to no work, some of them require potentially expensive processing. The fastest case is when there is no character needing percent encoding, no percent character, no backslash, and no dot character. In this fastest case, we may copy the path string as is. The second fastest case is when we can do without percent-encoding and there are no backslash nor percent characters. In this second case, we may still need to do some processing (e.g., convert \texttt{/././} to \texttt{/}), but it is still relatively simple. These two cases amount to the  majority of path strings found in realistic scenarios. Finally, we fall back on the complete case where there might be backslashes and character encoding. The general case includes cases such as \texttt{http://www.google.com/path/\%2e./} which must be normalized to \texttt{http://www.google.com/} (maybe surprisingly) because \texttt{\%2e.} is equivalent to `\texttt{..}' which instructions us to shorten the path.
To quickly classify the path strings, we use the algorithm of Fig.~\ref{lst:path-signature-cpp} which identifies the type of characters present:
\begin{itemize}
    \item The first bit is set whenever there is a forbidden character that needs percent-encoding.
    \item The second bit is set whenever the backslash character is present.
    \item The third bit is set whenever a dot character is present.
    \item The fourth bit is set whenever the percent character is present.
\end{itemize}
We call the result of the function a path signature.
We could use SIMD instructions for the computation of the path signature---it would be beneficial for long paths---but our signature routine is already efficient.

\begin{figure}
\begin{tabular}{c} 
\begin{lstlisting}[style=customcpp]

static uint8_t path_signature_table[256] = {
    1, 1, 1, 1, 1, 1, 1, 1, 1, 1, 1, 1, 1, 1, 1, 1, 1, 1, 1, 1, 1, 1, 1, 1,
    1, 1, 1, 1, 1, 1, 1, 1, 1, 0, 1, 1, 0, 8, 0, 0, 0, 0, 0, 0, 0, 0, 4, 0,
    0, 0, 0, 0, 0, 0, 0, 0, 0, 0, 0, 0, 1, 0, 1, 1, 0, 0, 0, 0, 0, 0, 0, 0,
    0, 0, 0, 0, 0, 0, 0, 0, 0, 0, 0, 0, 0, 0, 0, 0, 0, 0, 0, 0, 2, 0, 0, 0,
    1, 0, 0, 0, 0, 0, 0, 0, 0, 0, 0, 0, 0, 0, 0, 0, 0, 0, 0, 0, 0, 0, 0, 0,
    0, 0, 0, 1, 0, 1, 0, 1, 1, 1, 1, 1, 1, 1, 1, 1, 1, 1, 1, 1, 1, 1, 1, 1,
    1, 1, 1, 1, 1, 1, 1, 1, 1, 1, 1, 1, 1, 1, 1, 1, 1, 1, 1, 1, 1, 1, 1, 1,
    1, 1, 1, 1, 1, 1, 1, 1, 1, 1, 1, 1, 1, 1, 1, 1, 1, 1, 1, 1, 1, 1, 1, 1,
    1, 1, 1, 1, 1, 1, 1, 1, 1, 1, 1, 1, 1, 1, 1, 1, 1, 1, 1, 1, 1, 1, 1, 1,
    1, 1, 1, 1, 1, 1, 1, 1, 1, 1, 1, 1, 1, 1, 1, 1, 1, 1, 1, 1, 1, 1, 1, 1,
    1, 1, 1, 1, 1, 1, 1, 1, 1, 1, 1, 1, 1, 1, 1, 1};

uint8_t path_signature(std::string_view input) {
  uint8_t accumulator{};
  for (size_t i = 0; i < input.size(); i++) {
    accumulator |= path_signature_table[uint8_t(input[i])];
  }
  return accumulator;
}
\end{lstlisting}
  \end{tabular}
  \caption{Path-signature function}\label{lst:path-signature-cpp}
\end{figure}









\begin{figure}
    \centering

    \includegraphics[trim={5cm 2cm 1cm 1cm},width=0.50\textwidth]{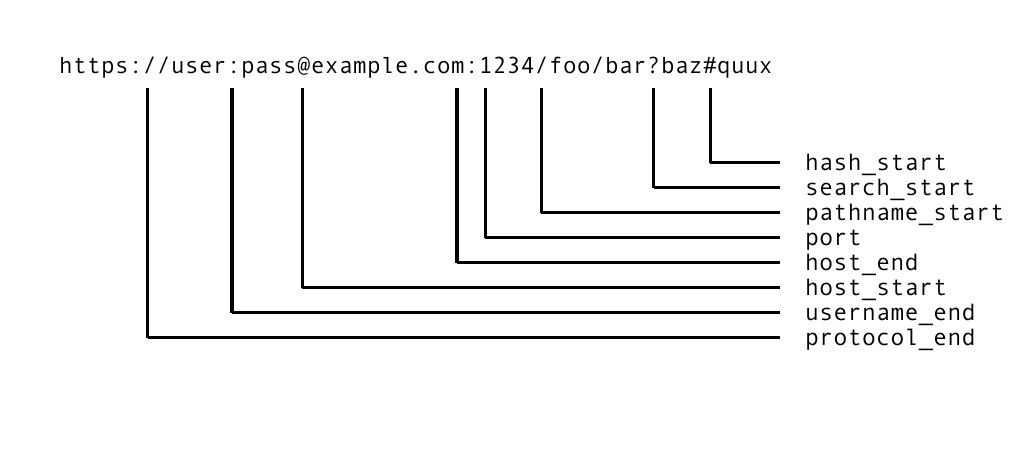}
    \caption{Component indexes}
    \label{fig:my_components}
\end{figure}

As we parse the input strings, we store the components (e.g., protocol, hostname) on a single buffer that becomes our normalized string. To record the location of the components, we use a convention similar to other parsers (e.g., Servo rust-url): counting the normalized string length, we only need nine~integers to characterize a parsed URL\@. See Fig.~\ref{fig:my_components}. In our actual implementation compiled with GCC~12 under Linux, we use 80~bytes per URL (not counting the dynamic memory allocation), of which 32~bytes are used by the \texttt{std::string} instance that we use as our buffer. Though our memory usage could be further optimized, it is clear that storing multiple 
\texttt{std::string} instances would use much more memory.

\subsection{JavaScript Integration}

In a system like Node.js, calling C++ from JavaScript can be relatively expensive. Indeed, creating a new JavaScript string instance from C++ data can be a costly operation. With our design where we have a single normalized string, we just need to additionally pass some integer offsets to indicate the position of the components in the string.  We also provide JavaScript with the protocol type as an integer, which allows (for example) to check that we have a file URL with a single integer comparison.
Components such as protocol, hostname, pathname, search, hash, etc.\ are computed as needed as substrings of the normalized string from within JavaScript.
In effect, we reduce as much as possible the need to copy strings between C++ and JavaScript, relying instead on  integer values.

\section{Benchmarks}

For C++ benchmarking, we use the release 2.4.1 for the Ada~library. Our implementation is \emph{safe} and \emph{correct} in the sense that it has undergone thorough testing,  including extensive tests with random inputs (fuzzing).

To directly compare our C++ implementation, we use the following competitors:
\begin{itemize}
    \item A high-quality WHATWG URL C++ library published as open-source software by Misevičius.\footnote{\url{https://github.com/rmisev/url_whatwg}}  We use a code snapshot from January~26 2023.
    \item We use the  Boost.URL C++ library version~1.81.0. 
    \item We use the rust-url library (version 0.1.0) from the Servo web browser engine\footnote{\url{https://servo.org}}, building it with Rust~1.65. The Firefox browser relies on the rust-url library.
\item We also use curl~7.81.0.
\end{itemize}
Both curl and Boost.URL follow the RFC~3986 standard~\cite{rfc3986} so direct comparisons must be done with care. We believe that the WHATWG URL standard is more demanding: we expect that all  RFC~3986 URLs are valid WHATWG URLs.

We considered adding URL parsers from major browsers (Chrome, Safari, etc.), but we were not able to use them as standalone components. Some of the browsers rely on customized memory allocators and other specialized code that is difficult to remove or isolate.
We found other URL parsers, but we believe that the standalone parsers we have selected are representative of the state-of-the-art: all of them are well maintained, reasonably fast, and well documented.

Our benchmark code consumes the URLs taken from large datasets: we ask each parser to normalize the strings. We use Google Benchmarks to derive accurate timings. We also add additional code to capture CPU performance counters (cycles and instructions retired).

\begin{figure}\centering
 \begin{subfigure}[h]{0.49\textwidth}
 \includegraphics[width=0.99\textwidth]{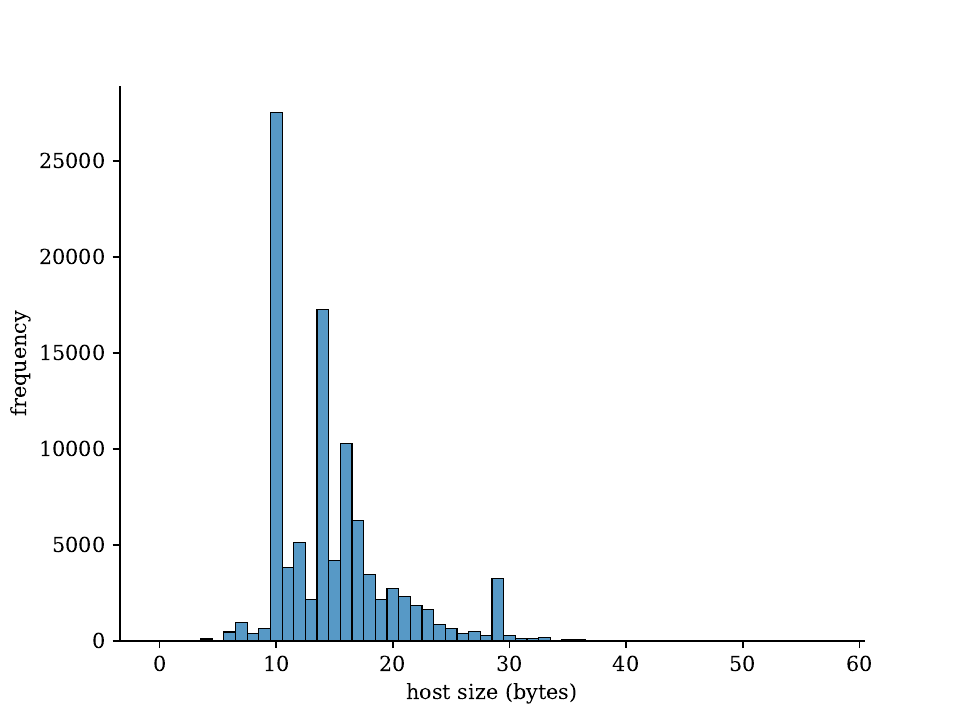}
\caption{host-size histogram} \end{subfigure}
 \begin{subfigure}[h]{0.49\textwidth}
 \includegraphics[width=0.99\textwidth]{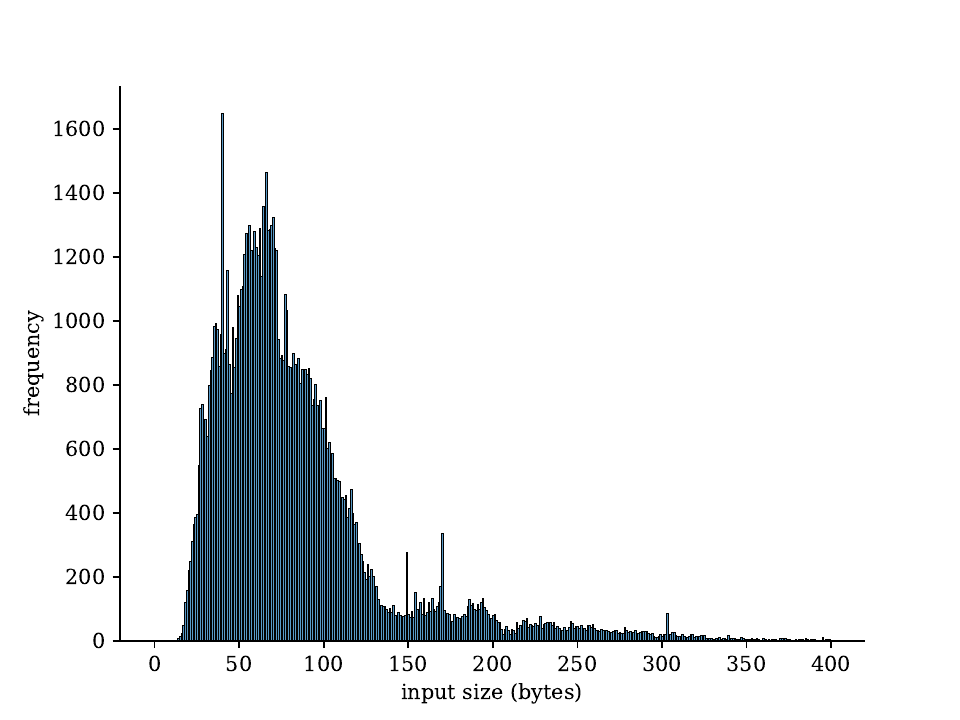}
\caption{total-size histogram)} \end{subfigure}
\caption{\label{fig:histograms}Histograms for the top~100 URL dataset (\num{100031}~URLs)}
\end{figure}

We gathered a collection of realistic URLs for benchmarking purposes and we make them freely available.\footnote{\url{https://github.com/ada-url/url-various-datasets}}
\begin{itemize}
\item The wikipedia~100k dataset contains \num{100000}~URLs from a snapshot of all Wikipedia articles as URLs (collected March 6th, 2023).
\item The top~100 dataset contains \num{100031}~URLs found in a crawl of the top~100 most popular websites on the Internet. It contains some invalid URLs: 
26~URLs according to the WHATWG URL specification are invalid. The curl parser finds 130~invalid URLs whereas the Boost.URL parser identifies 201~invalid URLs. We make freely available the JavaScript software we used to construct this dataset.\footnote{\url{https://github.com/ada-url/url-dataset}} Fig.~\ref{fig:histograms} presents two histograms regarding this dataset. The first histogram shows that the size of the host in bytes ranges roughly between 10 and 30 bytes, with some outliers. The total size of URL string ranges between a few bytes and hundreds of bytes. Most URLs use between 50~and 100~bytes.
\item  The Linux~files  dataset contains all files from a Linux system as URLs (\num{169312}~URLs).
\item The userbait dataset contains \num{11430}~URLs from a phishing benchmark.\footnote{\url{https://github.com/userbait/phishing_sites_detector}}
\end{itemize}
In some experiments, we also include another dataset: the kasztp dataset is made of \num{48009}~URLs from a URL shortener benchmark.\footnote{\url{https://github.com/kasztp/URL_Shortener}}

When they are not ASCII, all URLs are processed as UTF-8 strings. The conversion from UTF-16 inputs to UTF-8 would be take little computation~\cite{lemire2022transcoding}. We assume that all inputs are valid Unicode, validation would be similarly require little computation~\cite{keiser2021validating}.


 We run our benchmarks on the two systems presented in Table~\ref{tab:test-cpus}. The AMD server runs Ubuntu 22.04 whereas the Apple processor is on a standard MacBook Air (2022). We monitor the effective frequency and find that the MacBook Air remains at \SI{3.0}{\GHz} whereas the AMD servers maintain \SI{3.4}{\GHz}. We find little variation in the effective frequency between tests (within \SI{1}{\percent}).
 Our benchmark should not be interpreted as an assessment of the performance of ARM versus x64, or of AMD versus Apple. We use different hardware systems, released at different times, to arrive at a more robust comparison of the software.

\begin{table}
\caption{\label{tab:test-cpus} Systems 
}
\centering
\begin{minipage}{\textwidth}
\centering
\begin{tabular}{cccccc}\toprule
Processor   &  Frequency  & Microarchitecture                           & Memory  & Compiler\\ \midrule
AMD EPYC 7262   & \SI{3.4}{\GHz}  & Zen~2 (x64, 2019) &  DDR4 (3200\,MT/s)  & GCC  11 \\
Apple M2  & \SI{3.0}{\GHz}  & Avalanche (aarch64, 2022) &  LPDDR5 (6400\,MT/s)  & Apple/LLVM  14 \\ 
\bottomrule
\end{tabular}
\end{minipage} 
\end{table}

\begin{figure}\centering
 \begin{subfigure}[h]{0.49\textwidth}
 \includegraphics[width=0.99\textwidth]{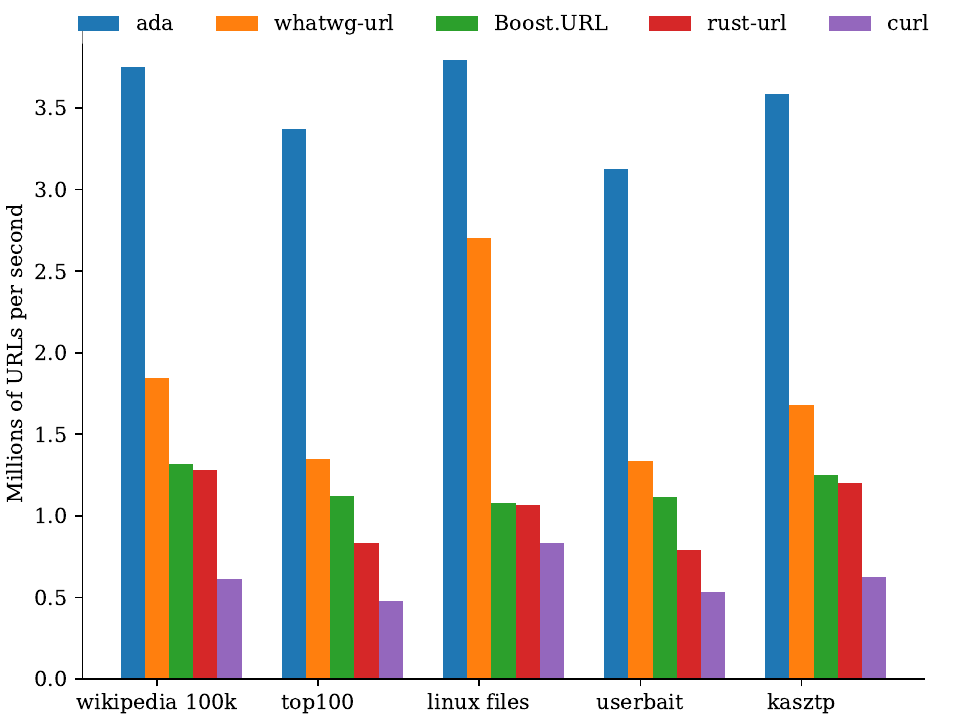}
\caption{AMD Rome (x64)} \end{subfigure}
 \begin{subfigure}[h]{0.49\textwidth}
 \includegraphics[width=0.99\textwidth]{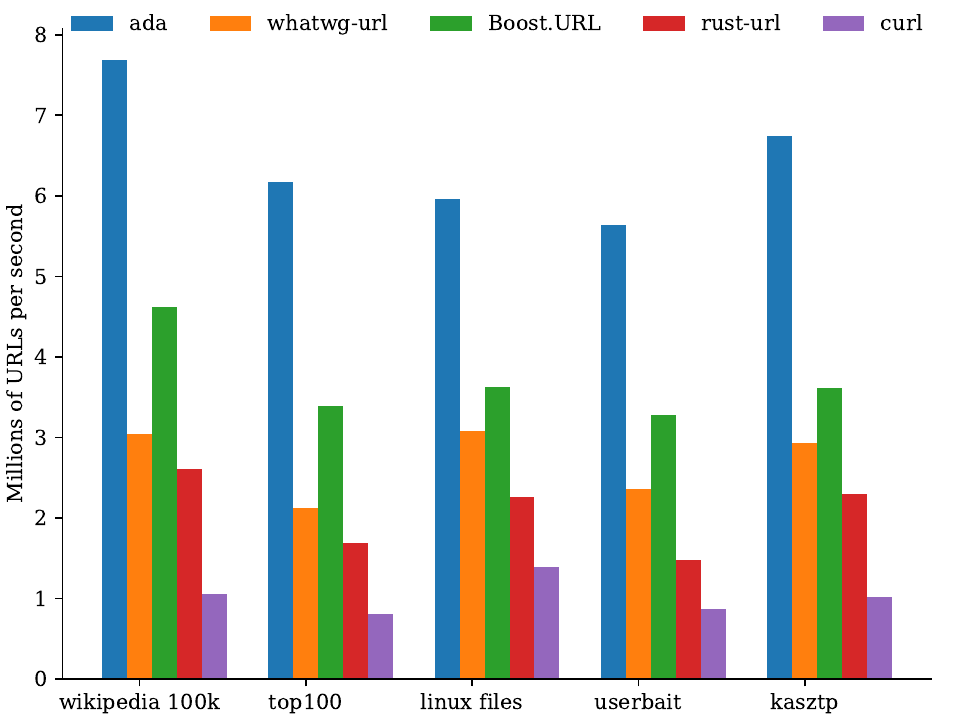}
\caption{Apple M2 (ARM64)} \end{subfigure}
\caption{Millions of URLs processed per second in C++ benchmarks\label{fig:speedcpp}}
\end{figure}

Fig.~\ref{fig:speedcpp} gives the number of millions of URLs processed per second for different datasets and different software libraries. Our parser (Ada) dominates, being often twice as fast as other parsers.
It is consistently faster than 3~million~URLs per second
on the AMD system,
and faster than 5~million~URLs per second
on the Apple system.
On the Linux~files dataset, the WHATWG~URL C++ parser has excellent performance on the AMD system, exceeding 2.5~million~URLs per second. The curl parser is the slowest in our tests: its performance is approximately 0.5~million~URLs per second on the AMD system, and nearly 1~million~URLs per second on the Apple system.

Table~\ref{table:rome} presents the collected performance counters while running the C++ benchmark, while Table~\ref{table:apple} has the performance counters for the Apple system. Ada requires consistently fewer instructions than the other parsers.  For example, on the top~100 dataset, it required  2200~instructions per URL (AMD) and 2400~instructions per URL (Apple) compared to \num{18000} and \num{19000} for curl: Ada required eight times fewer instructions.

\begin{table}\footnotesize\centering
\caption{Performance counters for Rome system\label{table:rome}}
    \begin{subtable}[h]{0.49\textwidth}
\caption{wikipedia 100k}
\begin{tabular}{lccc}
\toprule
name & instr./URL & cycles/URL & instr./cycle \\\midrule
ada & 2000 & 910 & 2.2 \\
WHATWG~URL & 4800 & 1800 & 2.7 \\
Boost.URL & 6200 & 2600 & 2.4 \\
rust-url & 6600 & 2600 & 2.5 \\
curl & 14000 & 5500 & 2.5 \\
\bottomrule
\end{tabular}
 \end{subtable}
 \restartrowcolors{}
    \begin{subtable}[h]{0.49\textwidth}
\caption{top 100}
\begin{tabular}{lccc}
\toprule
name & instr./URL & cycles/URL & instr./cycle \\\midrule
ada & 2200 & 1000 & 2.2 \\
WHATWG~URL & 6700 & 2500 & 2.7 \\
Boost.URL & 7200 & 3000 & 2.4 \\
rust-url & 9500 & 4000 & 2.4 \\
curl & 18000 & 7100 & 2.5 \\
\bottomrule
\end{tabular}
 \end{subtable}
  \restartrowcolors{}
    \begin{subtable}[h]{0.49\textwidth}
\caption{Linux files}
\begin{tabular}{lccc}
\toprule
name & instr./URL & cycles/URL & instr./cycle \\\midrule
ada & 2000 & 890 & 2.2 \\
WHATWG~URL & 3700 & 1200 & 2.9 \\
Boost.URL & 7600 & 3100 & 2.4 \\
rust-url & 7400 & 3200 & 2.3 \\
curl & 11000 & 4100 & 2.6 \\
\bottomrule
\end{tabular}
 \end{subtable}
  \restartrowcolors{}
    \begin{subtable}[h]{0.49\textwidth}
\caption{userbait}
\begin{tabular}{lccc}
\toprule
name & instr./URL & cycles/URL & instr./cycle \\\midrule
ada & 2100 & 1100 & 2.0 \\
WHATWG~URL & 5600 & 2500 & 2.3 \\
Boost.URL & 6500 & 3000 & 2.2 \\
rust-url & 9400 & 4300 & 2.2 \\
curl & 15000 & 6300 & 2.4 \\
\bottomrule
\end{tabular}
 \end{subtable}
  \restartrowcolors{}
 \end{table}

\begin{table}\footnotesize\centering
\caption{Performance counters for Apple~M2 system\label{table:apple}}
    \begin{subtable}[h]{0.49\textwidth}
\caption{wikipedia 100k}
\begin{tabular}{lccc}
\toprule
name & instr./URL & cycles/URL & instr./cycle \\\midrule
ada & 2000 & 440 & 4.6 \\
WHATWG~URL & 5900 & 1100 & 5.3 \\
Boost.URL & 3600 & 740 & 4.9 \\
rust-url & 7200 & 1300 & 5.5 \\
curl & 15000 & 3200 & 4.5 \\
\bottomrule
\end{tabular}
 \end{subtable}
 \restartrowcolors{}
    \begin{subtable}[h]{0.49\textwidth}
\caption{top 100}
\begin{tabular}{lccc}
\toprule
name & instr./URL & cycles/URL & instr./cycle \\\midrule
ada & 2400 & 550 & 4.5 \\
WHATWG~URL & 8200 & 1600 & 5.1 \\
Boost.URL & 4500 & 1000 & 4.5 \\
rust-url & 10000 & 2000 & 5.0 \\
curl & 19000 & 4200 & 4.5 \\
\end{tabular}
 \end{subtable}
  \restartrowcolors{}
    \begin{subtable}[h]{0.49\textwidth}
\caption{Linux files}
\begin{tabular}{lccc}
\toprule
name & instr./URL & cycles/URL & instr./cycle \\\midrule
ada & 2400 & 570 & 4.2 \\
WHATWG~URL & 5600 & 1100 & 5.1 \\
Boost.URL & 4100 & 920 & 4.4 \\
rust-url & 7900 & 1500 & 5.2 \\
curl & 12000 & 2500 & 4.7 \\
\bottomrule
\end{tabular}
 \end{subtable}
  \restartrowcolors{}
    \begin{subtable}[h]{0.49\textwidth}
\caption{userbait}
\begin{tabular}{lccc}
\toprule
name & instr./URL & cycles/URL & instr./cycle \\\midrule
ada & 2200 & 600 & 3.7 \\
WHATWG~URL & 6700 & 1400 & 4.7 \\
Boost.URL & 4200 & 1000 & 4.1 \\
rust-url & 10000 & 2300 & 4.5 \\
curl & 16000 & 3900 & 4.1 \\
\bottomrule
\end{tabular}
 \end{subtable}
  \restartrowcolors{}
 \end{table}

\subsection{JavaScript runtime environments}

We also benchmark URL parsing within JavaScript runtime environments. We used Node.js which can run JavaScript on servers using the Google v8 JavaScript engine. It contains additional code written in C++ and JavaScript.
Apart from the popular Node.js runtime environment, we selected two similar environments. Deno resembles Node.js in that it also relies on the v8 JavaScript engine; it is written in Rust instead of C++. Bun is another JavaScript environment but it replaces Google v8 with the WebKit's JavaScript engine (upon which Apple Safari is based). Bun is also written in part with C++ and Zig. For URL parsing, Bun relies on WebKit's C++ internal code whereas Deno uses rust-url\footnote{\url{https://github.com/servo/rust-url}}.
We use Deno (version~1.32.5), Bun (version~0.5.9), and Node.js (versions 18.15.0 and 20.1.0). We choose Node.js version 18.15 because more recent versions of Node.js include some of our URL-parsing software.
All systems run the same scripts, parsing the URLs from the top~100 dataset. We use mitata (version~0.1.6) as the benchmarking framework in JavaScript.\footnote{\url{https://github.com/evanwashere/mitata}}
We make our script available.\footnote{\url{https://github.com/ada-url/js_url_benchmark}}





\begin{figure}\centering
 \begin{subfigure}[h]{0.49\textwidth}
 \includegraphics[width=0.99\textwidth]{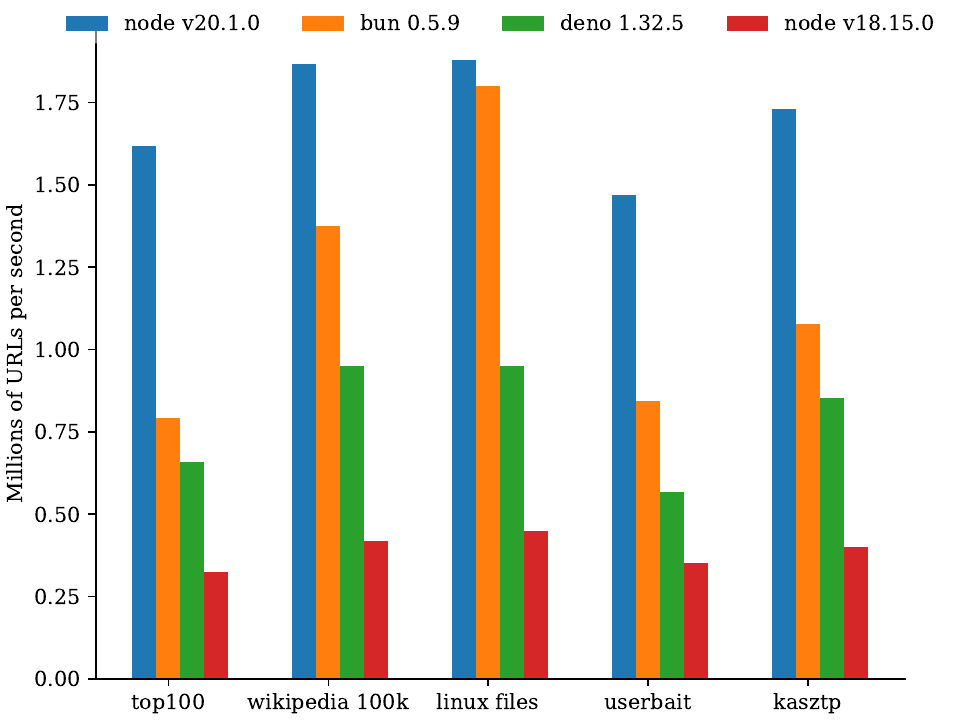}
\caption{AMD Rome (x64)} \end{subfigure}
 \begin{subfigure}[h]{0.49\textwidth}
 \includegraphics[width=0.99\textwidth]{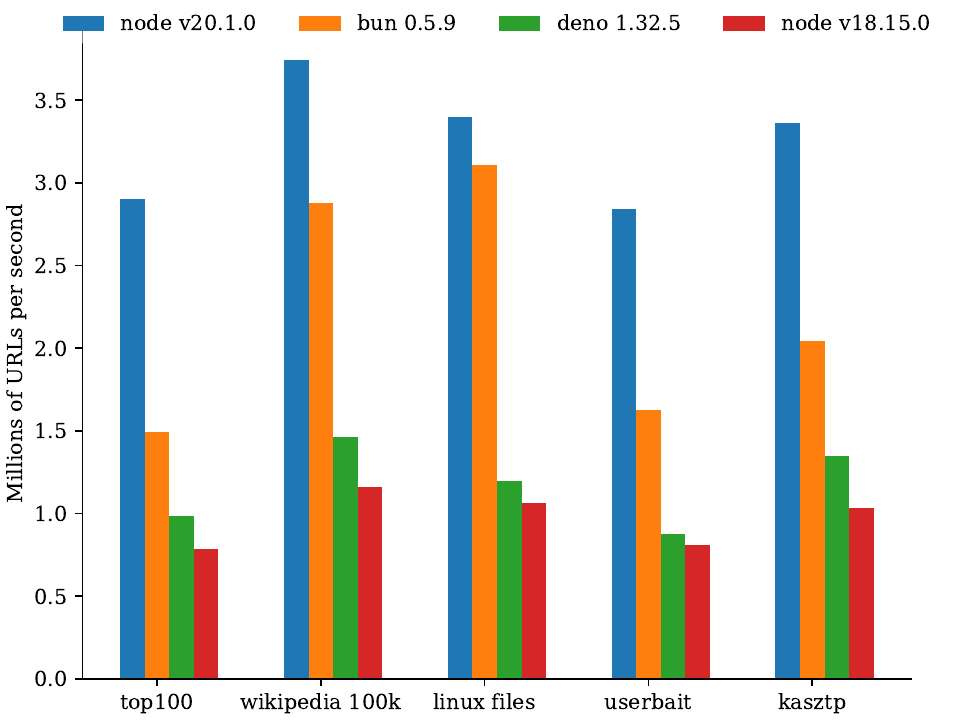}
\caption{Apple M2 (ARM64)} \end{subfigure}
\caption{Millions of URLs processed per second in JavaScript runtime environments\label{fig:speedjavascript}}
\end{figure}

Fig.~\ref{fig:speedjavascript} gives the number of millions of URLs processed per second for different datasets and different JavaScript systems. Node.js~20, with our Ada URL parser has the best performance. However, Bun also provides excellent performance, especially on the linux~files dataset where it comes close to Node.js~20. Roughly speaking, compared to the C++ benchmarks (Fig.~\ref{fig:speedcpp}), the speeds are about half: Node.js~20 is consistently faster than 
1.5~million~URLs per second on the AMD system,
and faster than 2.5~million~URLs per second on the Apple system. It suggests that about half of the processing is tied to the JavaScript system, some of it spent in C++, and the rest in JavaScript. The most important difference is between Node.js~20 and Node.js~18 (which lacked Ada): Node.js~20 is  four times faster on the Apple system and five times faster on the AMD system. We believe that it is essentially attributable to the replacement of the legacy URL parser by Ada. Node.js went from having the worst performance on URL parsing to the best performance compared to Bun and Deno. 

To gain further evidence that the better performance in Node.js is largely due to our work, we used profiling. Specifically, we ran the Node.js benchmarks using version~20 (with Ada) and version~18 (without Ada)
under the Linux \texttt{perf record} command. The command gathers
profiling data of a Linux application at \SI{4000}{\hertz} (by default). We then used the \texttt{perf report} command to identify the most time-consuming functions.
For Node~18, we finnd that the most time-consuming function related to URL parsing is
\texttt{node::url::URL::Parse}: it takes an estimated \SI{4.7}{\second} during the entire benchmark (all data files included). For Node~20, we find that the most time-consuming function related to URL parsing is
\texttt{ada::parser::parse\_url<ada::url\_aggregator>}. It takes an estimated \SI{1.4}{\second}, so between three and four times less time than the equivalent function in Node~18.

Node.js~20 is more than twice as fast as Deno in our experiments: it is consistent with the fact that Node.js~20 relies on the ada C++ library whereas Deno relies on rust-url, a significantly slower software library (see Fig.~\ref{fig:speedcpp}).

%

\subsection{JavaScript server}

To better assess the real-world performance impact of our URL parser, we wrote an http server (Fig.~\ref{lst:server}). The server supports two different paths: 
\begin{itemize}
    \item \texttt{simple}  returns the string contained in the body of the query;
        \item \texttt{href} parses the string as a URL and returns its normalized form.
\end{itemize}
We run the server locally using Node.js and benchmark its performance with the \emph{autocannon} http benchmarking tool\footnote{\url{https://github.com/mcollina/autocannon}}: multiple requests are issued during 10~seconds, using 10~threads.
We use a test query the JSON document \texttt{\{ "url": "https://www.google.com/hello-world?query=search\#value" \}}. 
We present the results from the Apple system in Table~\ref{table:httpapple}. The margin of error on the average number of requests per second is small (1\%) during our tests. 
 Our results suggest that Node.js~20.1 might be slightly faster than Node.js~18.15 on trivial requests (\texttt{simple}) by up to 2\%. However,  we find that Node.js~20.1 is faster by  1 $\approx 10$\%  compared to Node.js~18.15 for the requests that involve the URL parsing (\texttt{href}): it suggests that URL parsing could be a performance bottleneck in Node.js~18.15---prior to the integration of our URL parsing library.  We find interesting that when using Node.js~20, there is little difference between the \texttt{simple} and the \texttt{href} benchmarks ($\approx 3$\%) which suggests that URL parsing may no longer be a performance bottleneck.

\begin{figure}
\begin{tabular}{c} 
\begin{lstlisting}[style=customcpp]
const f = require('fastify')()

f.post('/simple', async (request) => {
    const { url } = request.body
    return { parsed: url }
})

f.post('/href', async (request) => {
    const { url } = request.body
    return { parsed: new URL(url).href }
})

f.listen({ port: 3000 })
    .then(() => console.log('listening on port 3000'))
    .catch(err => console.error(err))
\end{lstlisting}
  \end{tabular}
  \caption{Node.js http server.}\label{lst:server}
\end{figure}

\begin{table}\footnotesize\centering
\caption{http performance for Apple~M2 system\label{table:httpapple}}
\begin{tabular}{rcc}
\toprule
node~version & request/second (simple) &  request/second (href)\\\midrule
20.1 & 61k & 59k \\
18.15 & 60k & 54k \\
\bottomrule
\end{tabular}
  \restartrowcolors{}

 \end{table}

\section{Conclusion}

We developed and released a new URL parser
that provides full compliance with the WHATWG URL specification.
It replaced the legacy Node.js parser, multiplying the performance of URL parsing in Node.js.
We believe that our good results can be explained in large part by the following  strategies: (1)~reduce the number of memory allocations to a minimum, using a single buffer if possible, (2)~implement fast functions to check for common fast paths, (3)~replace strings with simpler types such as integers whenever possible. Our work suggests that there is still much room for performance 
improvements in the software used to build web applications.

We provide various language bindings for our C++ URL parsing library,\footnote{\url{https://github.com/ada-url}} including C, Rust, Python, Go. Our Rust binding has a performance similar to the C++ library, and  it is several times faster than the popular rust-url library. 
Future work could examine how to improve the performance of URL parsing in other important systems (e.g., Deno), possibly by using our software library and its bindings. 

Future work could consider more advanced techniques. For example,
we could design single-instruction-multiple-data (SIMD) algorithms
able to benefit from the powerful new instruction sets (e.g., AVX-512, SVE2).
We expect that significant gains in URL parsing are still possible.
There are other important components of modern web applications that could be optimized.

\section*{Author Contributions}

Yagiz Nizipli: conceptualization; investigation; software; Node.js integration; experimentation; writing-review and editing. 
Daniel Lemire: conceptualization; software; validation; experimentation; data analysis; writing-original draft; writing-review and editing.

\section*{Acknowledgements}

We thank A.~Henningsen, M.~Teixeira, R.~Nagy, S.~Klabnik,
N.~Nuon, S.~Vohr, M.~Atlow, and D.~Chatterjee for their software contributions.
We thank A.~van Kesteren for answering our technical questions on the URL standard.
We thank that Node.js team for their feedback.

\section*{Data Availability Statement}

All our data and software is freely available online. The C++ benchmarking software is available online
at \url{https://github.com/ada-url/ada}.
We make the URL datasets available online at \url{https://github.com/ada-url/url-various-datasets}. We wrote and published a URL-parsing JavaScript benchmark which is available online at \url{https://github.com/ada-url/js_url_benchmark}.
We collected performance data and processed it using Python scripts: both the raw data and the scripts are available online at \url{https://github.com/ada-url/ada_analysis}. We also published an http server for benchmarking purposes at \url{https://github.com/ada-url/http-benchmark}.



\bibliography{urlparser}


\end{document}